\documentclass[aps,prl,twocolumn,superscriptaddress,showpacs]{revtex4}
\usepackage{graphicx}
\usepackage{bm}
\usepackage{color}
\input colordvi
\begin{document}
\title{Nonequilibrium quantum dynamics of a charge carrier doped into a Mott insulator}
\author{M. Mierzejewski}
\affiliation{Institute of Physics, University of Silesia, 40-007 Katowice, Poland}
\affiliation{ J. Stefan Institute, SI-1000 Ljubljana, Slovenia }
\author{L. Vidmar}
\affiliation{ J. Stefan Institute, SI-1000 Ljubljana, Slovenia }
\author{J. Bon\v{c}a}
\affiliation{ J. Stefan Institute, SI-1000 Ljubljana, Slovenia }
\affiliation{Department of Physics, FMF, University of Ljubljana, Jadranska 19, SI-1000 Ljubljana, Slovenia }
\author{P. Prelov\v{s}ek}
\affiliation{ J. Stefan Institute, SI-1000 Ljubljana, Slovenia }
\affiliation{Department of Physics, FMF, University of Ljubljana, Jadranska 19, SI-1000 Ljubljana, Slovenia }
\begin{abstract}
We study real-time dynamics of a charge carrier introduced into undoped Mott insulator propagating  under a constant electric field  $F$ on the  $t$--$J$ ladder and  square lattice.
We calculate quasistationary current. In both systems adiabatic regime is observed followed by the positive differential resistivity (PDR)  at moderate  fields where carrier  mobility is determined. Quantitative differences between ladder and 2-dimensional (2D) system emerge when at large fields both systems enter negative differential resistivity (NDR) regime. In the ladder system Bloch-like oscillations prevail, while  in 2D  the  current remains finite, proportional to $1/F$. The crossover between PDR and NDR regime in 2D is accompanied  by  a change of the spatial structure of the  propagating spin polaron.
\end{abstract}

\maketitle

\textit{Introduction.}---
The real--time response of interacting many--body quantum systems remains in many aspects an unexplored field  that has recently attracted significant attention.
Understanding of the time--dependent
phenomena becomes vital for various branches of physics including condensed matter \cite{bulk_e},  
nanostructures  \cite{nano_e} and optical lattice systems \cite{opt_e,demler}.         
Since only very few cases are exactly solvable,
the vast majority of unbiased results has been obtained from 
numerical approaches like  exact diagonalization (ED) \cite{ED,fehske}, 
time--dependent density matrix renormalization group \cite{DMRG} 
or nonequilibrium dynamical mean--field theory \cite{DMFT}. 
Among others, the electric--field induced breakdown of the Mott insulator { (MI)} \cite{ED,tokura,mt1,eckstein}, 
nonequilibrium transport in nanostructures \cite{qd1,hmeis} and the relaxation of 
correlated systems  after the photoexcitations \cite{matsueda,koshibae}, represent the well--studied examples 
of nonequilibrium phenomena which are important both for fundamental understanding of strongly correlated systems as well as for their potential 
applications.

Most of theoretical studies so far considered the breakdown of {\em undoped} { MI}, 
when threshold  value of the electric field exceeds experimental value \cite{tokura} by
a few orders of magnitude (see discussion in Ref. \cite{eckstein}). 
It  indicates that other transport mechanism becomes active at energies much lower 
than the Mott--Hubbard gap. In this Letter we inveqstigate a nonequilibrium 
response of a charge carrier doped into the insulator and driven by an uniform electric field $F$.
Understanding of this subject on one hand widens our knowledge of  a charge carrier doped into the antiferromagnetic (AFM) background \cite{tj,janez1}, on the other, it 
represents a fundamental 
problem of a quantum particle moving in a dissipative medium \cite{vidmar}.

Having in mind strongly correlated systems, we investigate the $t$--$J$ model 
where the particle driven by a constant electric field dissipates the energy by inelastic scattering on spin degrees of freedom.
We use numerical approaches to treat $t$--$J$ model at zero temperature  on two different system geometries, i.e., a ladder with periodic boundary conditions and an infinite 2--dimensional (2D) square lattice.

The most important finding of this Letter concerns the current--field characteristic, where the linear
part with a well defined mobility is followed by a strongly nonlinear one with a
negative differential resistivity. Numerical results reveal also adiabatic evolution
for very weak fields (below the linear regime) and strong Bloch oscillations (BO)  in the opposite limit of large $F$.
These qualitative results hold for both the geometries. The most prominent difference between the ladder and the 2D case emerges at large field where in the latter system due to different topology, allowing for transverse charge carrier motion, BO remain damped and the steady current decreases with field as $1/F$. 

\textit{Model}---We consider a charge carrier within the $t$--$J$ model 
threaded by a time--dependent magnetic flux:
\begin{equation}
H=-t_0 \sum_{\langle \mathbf{l}\mathbf{j} \rangle,\sigma} \left[ {\mathrm e}^{i \phi_{\mathbf{l}\mathbf{j}}(t)}\; \tilde{c}^{\dagger}_{\mathbf{l}, \sigma}
\tilde{c}_{\mathbf{j}, \sigma} +{\mathrm H.c.} \right]+J \sum_{\langle \mathbf{l}\mathbf{j} \rangle} \mathbf{S_l} \cdot \mathbf{S_j},
\label{hdef}
\end{equation}      
where $\tilde{c}_{\mathbf{j},\sigma}=c_{\mathbf{j},\sigma}(1-n_{\mathbf{j},-\sigma})$ is a projected fermion operator and $\langle \mathbf{l}\mathbf{j} \rangle$ denote nearest neighbors.
The constant electric field $F$ is switched on at time $t=0$ and is measured in units of $[t_0/e_0a]$, where $e_0$ is the unit charge and $a$ is the lattice distance. { We set $t_0=\hbar=e_0=a=1$.
For the ladder the charge current operator for $t>0$ reads}
\begin{equation}
\hat{I}= i t_0 \sum_{j,\sigma} \left(  {\mathrm e}^{-iFt}  \; 
\tilde{c}^{\dagger}_{j+\hat{x},\sigma}\tilde{c}_{j,\sigma} -{\mathrm H.c.} \right),
\label{jdef}
\end{equation}   
where $F$ acts along the ladder's leg and ${\phi}_{\mathbf{l}\mathbf{j}}(t)$ equals $-Ft$ and $0$ for hopping in $\hat{x}$-- and 
$\hat{y}$--direction, respectively.

At this stage it is instructive to recall a simple relation between the current $j(t)=\langle \hat{I}(t)\rangle$ and the total energy $E(t)=\langle H(t)\rangle$  \cite{tv},
$\dot{E}(t)=F I(t) $, 
which allows one to calculate the steady component of $j(t)$ 
\begin{equation}
\bar{j} =\lim_{t \to \infty}\frac{1}{t}\int_0^t {\mathrm d}\tau j(\tau)  =  \frac{\Delta E(t)}{t F},
\label{idc}
\end{equation}
where  $\Delta E(t)= E(t)-E(0)$.  

\textit{Ladder.}---The real--time response of a ladder with $L$ rungs is studied in the full Hilbert space by means of ED. Applying the Lanczos technique we have determined the initial ground state $|\Psi(t=0) \rangle$ at  $\mathbf{p}_0=(\pi/2,0)$ \cite{greiter}.
The time evolution of the initial state is calculated by step--vise change of the flux $\phi_{lj}(t)$ in small time increments $\delta t \ll 1,$ employing at each step Lanczos basis 
generating the evolution $|\Psi(t-\delta t) \rangle \to |\Psi(t) \rangle$ \cite{lantime,tv}.

\begin{figure}
\includegraphics[width=0.47\textwidth]{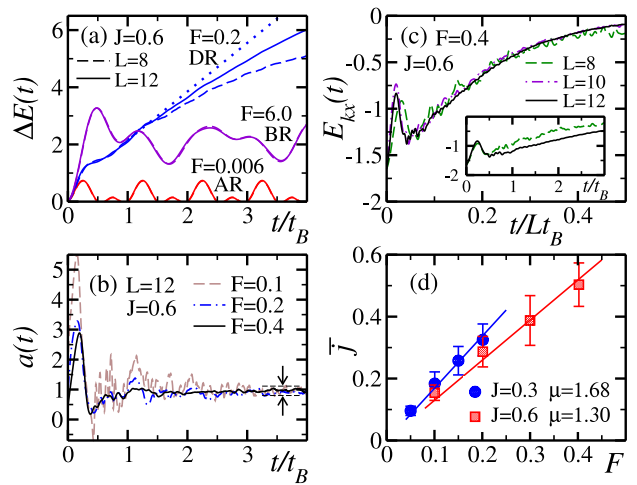}
\caption{
(Color online) Ladder with $L$ rungs: 
(a) increase of energy $\Delta E(t)$ in different regimes; 
(b) real--time current $j(t)$ normalized by kinetic energy $a(t)=j(t)/(F|E_{kx}(t)|)$ (note the steady $F$-independent ratio);
(c) kinetic energy $E_{kx}$ vs.  $t/t_{\mathrm B}$ (inset) and $t/Lt_{\mathrm B}$ (main);
(d) steady  current $\bar{j}$ vs. $F$. 
Dotted line in (a) shows $\Delta E(t)$ estimated from the mobility $\mu$ shown in (d) as linear fits 
$\bar{j}= \mu F$. 
$J=0.6$ is used in (a)-(c).}
\label{fig1}
\end{figure}

Fig. \ref{fig1}(a)  demonstrates how the total energy 
changes in time.
One can identify three regimes/limits of $F$: 
the adiabatic regime (AR) 
for $F\rightarrow 0$,  the Bloch--oscillations regime (BR) for $F \gg 1 $, and 
the dissipative regime (DR) for  intermediate $F$.  
In the adiabatic limit,  $E(t)$
follows the ground state dispersion $E_0(\mathbf{p})$ with $\mathbf{p}=\mathbf{p}_0+Ft\hat{x}$. 
In the ladder $E_0(\mathbf{p})$ is 
separated from excited states by a finite gap, related to spin gap. The dispersion is a quasiparticle (QP)  one, i.e. 
it follows the behavior of a single carrier with the periodicity $\Delta p_x=2\pi$. 
As a result $E(t)$  oscillates with the Bloch frequency $\omega_B=2 \pi /t_B=F$ also in AR. Due to finite gap, qualitatively similar behavior remains even for $|F|>0$.  We expect that the  crossover from AR to DR is connected with a transition through the gap 
resembling the Landau--Zener transition through the Mott--Hubbard gap \cite{ED}. { Hence, the threshold 
$F$ should be determined by $J$.}

Fig. \ref{fig1}(a) shows that
$E(t)$ is monotonic only in DR where $j(t)$ doesn't change sign. Other regimes (AR, BR) 
are characterized by strong  current oscillations. 
Fig. \ref{fig1}(a) together with Eq. (\ref{idc}) allow
one to estimate $\bar j$  as a function  of the electric field. Clearly $\bar j$ is 
{\em maximal} in DR,  therefore there  exists a corresponding  "optimal" $F$. 
Substantial differences between the regimes show up also in the kinetic energy related 
with the  movement along the $\hat{x}$--direction, $E_{kx}=\langle H_{kx} \rangle$ where 
$H_{kx}=-t_0 \sum_{j,\sigma} \left(  {\mathrm e}^{-iFt}  \; 
\tilde{c}^{\dagger}_{j+\hat{x},\sigma}\tilde{c}_{j,\sigma} +{\mathrm H.c.} \right)$.
We have found that $E_{kx}(t)$ oscillates in AR and BR. This quantity is
always negative in AR, whereas in BR it takes on negative as well as positive values.
Therefore, 
BR resembles the BO of noninteracting particles in that
both $j(t)$ and $E_{kx}(t)$ change sign. 

{ Nonzero $\bar{j}$ in DR} implies a steady growth of energy due to the Joule heating (although we are not dealing with well defined thermalization).
This effect is limited by the system size since a carrier driven by constant $F$ propagates many times around the ladder with finite $L$, thus steadily excites the spin background and increases its effective temperature. 
Fig. \ref{fig1}(a) shows that $\Delta E(t)$ deviates considerably 
between different $L$ at time $t_1\simeq 2t_B$.
As there is a single charge carrier,
the quantity $r(t)= \int_0^t {\mathrm d}\tau j(\tau)=\Delta E(t)/F $ [see Eq. (\ref{idc})] can be viewed as a distance traveled by a carrier within the time-interval $(0,t)$. 
For noninteracting case
$\Delta E(t) < 4$ and $r(t) < 4/F$, which is the Stark localization length \cite{kolovsky}.
Here, we have found that $r(t_1) \sim L$ { confirming}
that the finite--size effects originate from heating of the spin background when carrier
repeatedly encircles the ladder.

However, even for $t>t_1$ one can follow the behavior in a controlled 
way. 
Previous analysis of 1D system of spinless fermions 
has shown that  heating can be accounted for 
by a renormalization of the kinetic energy, as a sum rule for the optical conductivity $\sigma(\omega)$ \cite{tv}. 
Results in Fig. \ref{fig1}(b) show that the same reasoning can be applied to the $t$--$J$ ladder.
In particular, the ratio  $a(t)= j(t)/(F|E_{kx}(t)|)$ reaches already at short $t/t_B$ nearly a constant value that is independent of $F$ and $L$. 
Therefore, the original problem concerning the evaluation of $\bar{j}$ can be reduced to finding $E^*_{kx}=E_{kx}(t \to \infty)$ in the 
limit $L \to \infty$.

A straightforward expectation is that for
$n$--times larger system ($L \rightarrow n L$) it takes $n$ times longer
to reach the same average temperature of the spin  background.
It explains the $E_{kx} \propto  E_{kx}(t/L)$ scaling visible for $t > t_1$
in Fig. \ref{fig1}(c). The initial value of $E_{kx}$ in this time--domain
represents
the upper bound on $E^*_{kx}$  when $L\rightarrow \infty$.
On the other hand, for $t < t_1$ results for $E_{kx}(t)$ merge without any rescaling of
time [see the inset
in Fig. \ref{fig1}(c)]. Hence the final value of $E_{kx}$ in the latter
time--domain poses
the lower bound on $E^*_{kx}$.  These bounds on $E^*_{kx}$ together with
the ratio  $a(t)$
allow one to extract $\bar{j}$
{ (see Fig. \ref{fig1}(d))}. 
Within presented range of $F$, in the DR nonlinear effects are { weak} and one can easily estimate the hole mobility: $\mu\sim 1.7$ for $J=0.3$ and $\mu\sim 1.3$ for $J=0.6$. 
However, significant nonlinear effects have to show up for larger fields since $\bar{j}$ 
is maximal for a finite $F$. { This observation together with the value of $\mu$ are
the main results for the ladder system.}

\begin{figure}
\includegraphics[width=0.47\textwidth]{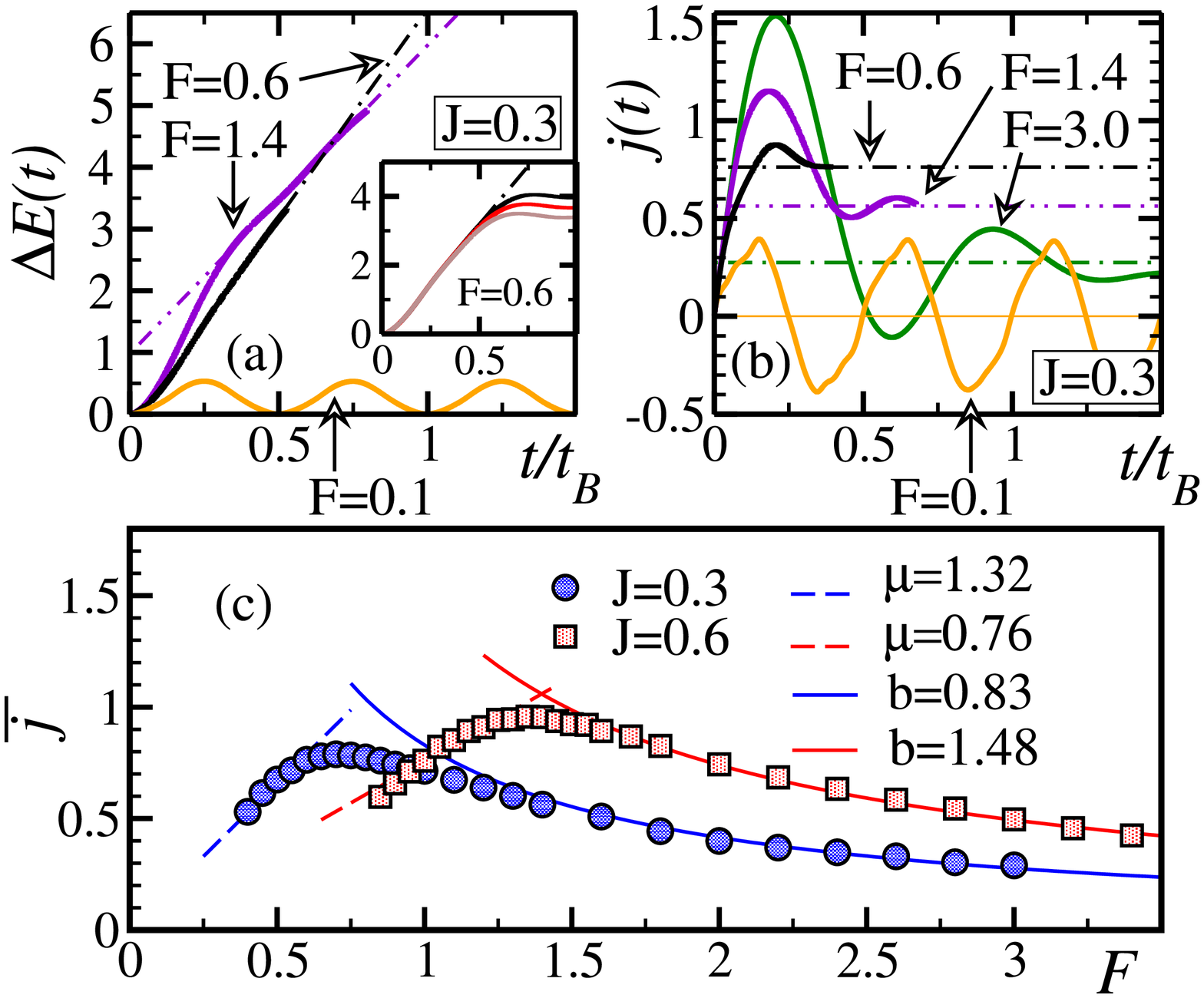}
\caption{
(Color online) 2D lattice:
(a) increase of energy $\Delta E(t)$  for $J=0.3$, various $F$ (main) and  $N_h=12,13,14$ (inset);
(b) real--time current $j(t)$, $\bar j$ from Eq.~\ref{idc} (dotted--dashed lines);
(c) steady current $\bar{j}$ vs. $F$ fitted by $\bar{j}=\mu F$ (dashed lines) and  $\bar{j}=b/F$ (solid lines) in PDR and NDR regimes, 
respectively. 
}
\label{fig2}
\end{figure}

\textit{2D square lattice.}---Since in the ground state a single carrier in 2D lattice carries momentum $\mathbf{k}_0=(\pi/2,\pi/2)$, we set ${\phi}_{\mathbf{l}\mathbf{j}}(t)=-Ft/\sqrt{2}$ for $\hat{x}$-- as well as $\hat{y}$--direction, and $\omega_B=F/\sqrt{2}$.
Accordingly, we also define and calculate the modified current $j(t)$, Eq. (\ref{jdef}), along the diagonal. 
We employ an ED method at defined over a limited functional space (EDLFS) which describes properties of a carrier doped into planar ordered  AFM  \cite{janez1}.
One starts from a translationally invariant state of a carrier in the N\' eel background $|\phi_0\rangle=c_{\mathbf{k}_0}|\mbox{N\' eel}\rangle$.
The kinetic part $H_k$ as well as the off--diagonal spin--flip part $\tilde{H}_J$ of the Hamiltonian (\ref{hdef}) are applied up to $N_h$ times generating the basis vectors:
$\left\{|\phi_{l}^{n_h} \rangle \right\}=
[H_k(F=0) + \tilde{H}_J]^{n_h}
|\phi_0 \rangle$ for $n_h=0,...,N_h$.
The ground state $|\Psi(t=0)\rangle$ and its time--evolution under electric field are calculated within the limited functional space in the same way as previously for the ladder within the full Hilbert space.
The advantage of EDLFS over the standard ED follows from systematic generation of selected states which contain spin excitations in the vicinity of the carrier. It  enables investigation of the
dynamics of large systems, which are far beyond the reach of ED.
Since $N_h$ determines the accessible energies (spin excitations), 
this quantity poses limits on the maximal  propagation time characterized by a steady growth of energy. 
We set $N_h=14$ while smaller values of $N_h$ are used to establish 
the time--window where results are independent of $N_h$ (see inset in Fig \ref{fig2}a).

AR at small $F=0.1$ is clearly seen from Figs. \ref{fig2}(a) and (b) that show $\Delta E(t)$ and $j(t)$ after the field has been switched on at $t=0$.  Both quantities are consistent with the adiabatic propagation along the QP band  with a period  $t_p=t_{\mathrm B}/2$, consistent with the AFM long range order. 
In the DR the carrier starts to move due to the constant field, it emits magnons and consequently, after propagation through the transient regime $t/t_B\ll 1$,  it develops a finite average velocity as seen for $F=0.6$ and 1.4. 
DR  is characterized by a constant current and a linear increase of the total energy of the system. 
To calculate the current  we make use of Eq. (\ref{idc}) where the linear increase of $\Delta E(t)$ provides the value of $\bar{j}$.
Indeed, linear dotted--dashed fits in Fig. \ref{fig2}(a) indicate that the system has already reached the quasi stationary  state.
We plot the extracted values of $\bar{j}$ as dotted--dashed horizontal lines in Fig. \ref{fig2}(b).
 Similar approach has recently been applied to the problem of a driven Holstein polaron \cite{vidmar}.
At large field, $F\gtrsim 3$, $j(t)$ is consistent with a damped BO with  $t_p\lesssim  t_{\mathrm B}$, signaling the onset of BR. 

\begin{figure}
\includegraphics[width=0.47\textwidth]{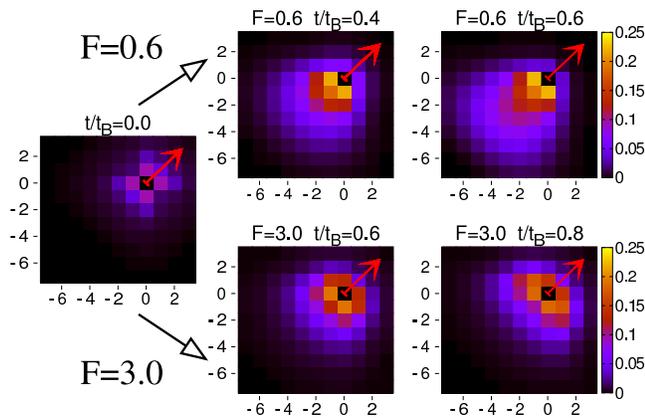}
\caption{
(Color online)
2D lattice with  $J=0.3$.
Snapshots of the spin deformation $C(\mathbf{r})$ in the vicinity of hole
[placed at $(0,0)$], taken at different times $t/t_{\mathrm B}$. 
Filled arrows indicate direction of the electric field.
}
\label{fig3}
\end{figure}

Fig. \ref{fig2}(c) displays $\bar j$--$F$ characteristics in DR for $J=0.3$ and $J=0.6$.
Each point has been calculated from Eq. (\ref{idc}) and then compared with $j(t)$, as demonstrated in Figs. \ref{fig2}(a) and (b).
The most remarkable property is a peak at $F_0$ dividing DR   into two sub-regimes, i.e., the regime
of positive  differential resistivity (PDR) for $F<F_0$ and the regime of negative differential resistivity (NDR) for $F>F_0$.
The value of the crossover field $F_0$ scales with  the exchange energy $J$. 
From curves in Fig. \ref{fig2}(c) we find $F_0 \simeq 2.3 J$.
This is consistent with an intuitive expectation that the onset of the NDR regime emerges when the energy gained by a single hop exceeds the maximal energy of one-magnon excitation.
Consequently, the real--space propagation of a carrier exhibits qualitatively different behavior in both regimes.
This is illustrated in Fig. \ref{fig3} which displays spin deformation function $C(\mathbf{r})$ around the carrier at different times.
We define $C(\mathbf{r}) = \sum_{\mathbf{i}} \langle n_{\mathbf{i}} \mid S_{\mathbf{i}+\mathbf{r}}^{z,\mbox{\tiny N\' eel}} - S_{\mathbf{i}+\mathbf{r}}^{z} \mid \rangle$ with
$S_{\mathbf{j}}^{z,\mbox{\tiny N\' eel}}=\pm \frac{1}{2}$.
In the PDR regime and for $F=0.6$ spin excitations predominantly emerge behind the traveling carrier indicating  that the average charge velocity $\bar v_c$ is larger than magnon velocity $v_s=\sqrt{2}J$, {\it i.e.} $\bar v_c=\bar j>v_s$ (see the upper panel of Fig. \ref{fig3} ).  A more complex pattern characterizes the NDR regime where at $F=3$, $\bar v_c<v_s$ (lower panel of Fig. \ref{fig3}) and enhanced spin excitations  develop also in front of the carrier. Moreover,  $C(\mathbf{r})$  displays a distinct  transverse orientation with respect  to the field direction.  This signals an increased transverse and longitudinal carrier oscillation  that serves  to release  the excess energy
to the unperturbed spin background. 


In the PDR regime (dashed lines in Fig. \ref{fig2}(c)) linear fits of the current increase provide an estimate for the carrier mobility, which yields $\mu\sim 1.3$ for $J=0.3$ and $\mu\sim 0.8$ for $J=0.6$.
These values are in agreement with ED calculations using the linear--response theory \cite{jaklic} and indicate that  carrier mobility increases with growing correlations.
In the NDR regime a scaling $\bar{j}\sim 1/F$ is found, Fig. \ref{fig2}(c).
This is consistent with incoherent hopping between Stark states appropriate for dispersive boson excitations \cite{emin}, in contrast to coherent hopping in the case of dispersionless phonons leading to $\bar{j}\propto 1/\sqrt{F}$ \cite{vidmar}.

\textit{Discussion.}---
At small $F$ the  ladder and  the 2D  system exhibit adiabatic propagation with $\bar j=0$. 
The validity of AR is
closely connected with the stability of the lowest QP band. 
While this is more evident for the ladder (due to a gap), 
the gapless (acoustic) magnons in 2D lattice
require more care. Nevertheless, our results show that the condition for 
Cerenkov radiation of magnons $v(k)> v_s = \sqrt{2} J$ is for $J\geq 0.3$ never fulfilled. In part, because  the QP bandwidth is limited 
by $J$ and not by $t$. For  $J=0.3$ we obtain maximal charge velocity $v_{\mathrm{max}}=1.36J$ while for $J=0.6$, $v_{\mathrm{max}}=1.10J$. Moreover, the QP weight remains finite throughout the whole Brillouin zone \cite{janez1,leung}.

With increasing $F$ both systems enter DR 
where  the quasistationary current is
proportional to $F$, leading to well defined mobility being
even quantitatively close for both cases. The mechanism of dissipation
is clearly the emission of spin excitations. Still the dissipation
is due to topological difference (in particular due to the possibility of
spin perturbation transverse to the direction of the carrier propagation) more efficient
within the 2D lattice  leading to stronger damping of the
BO and finite $\bar j\propto 1/F$.  Consequently, quasistationary
current on the square lattice
occurs up to  much larger $F$ than in the ladder where in the same regime
strong BO with $\bar j\sim 0$ prevail.

Surprisingly, the undoped \cite{eckstein} and lightly doped MI show 
the same sequence of the field--regimes:
$\bar{j}$ is zero or exponentially small for sufficiently weak $F$ (AR in the present case);
for larger $F$ the linear  $\bar{j}-F$ dependence is restored (DR); finally, for very large $F$
the response is dominated by the BO (BR). However, in doped and undoped MI 
these regimes occur at exceedingly  different fields. In particular,  the DMFT studies of the strong-$U$ limit \cite{eckstein} reveal that the threshold field for the dielectric breakdown of undoped MI is $F_{th}\gg 1$, which is an order of magnitude above the field discussed here.

\acknowledgements

L.V. and J.B. acknowledge stimulating discussions with T. Tohyama and O. P. Sushkov.
This work has been support by the Program P1-0044 of the Slovenian Research Agency (ARRS) 
and REIMEI project, JAEA, Japan.

\end{document}